\pdfoutput=1
\documentclass[%
reprint,
 amsmath,amssymb,
 aps,
prl
]{revtex4-1}

\usepackage{graphicx}
\usepackage{dcolumn}
\usepackage{bm}
\usepackage[dvips]{color}

\usepackage[utf8]{inputenc}
\newcommand{\be}{\begin{equation}}
\newcommand{\ee}{\end{equation}}
\newcommand{\ba}{\begin{eqnarray}}
\newcommand{\ea}{\end{eqnarray}}
\newcommand{\nn}{\nonumber}

\definecolor{red}{named}{Red}

\begin{document}
\title{Emulating quantum computation with artificial neural networks}
\author{Christian Pehle}
\email{christian.pehle@kip.uni-heidelberg.de}
\author{Karlheinz Meier}
\email{meierk@kip.uni-heidelberg.de}
\author{Markus Oberthaler}
\email{oberthaler@kip.uni-heidelberg.de}
\affiliation{%
Kirchhoff-Institute for Physics \\
Heidelberg University \\
Im Neuenheimer Feld 227, D-69120 Heidelberg
}%
\author{Christof Wetterich}
\email{wetterich@thphys.uni-heidelberg.de}
\affiliation{
Institute for Theoretical Physics \\
Heidelberg University \\
Philosophenweg 16, D-69120 Heidelberg
}%


\begin{abstract}
We demonstrate, that artificial neural networks (ANN) can be trained to emulate single or multiple basic quantum operations. In order to realize a quantum state, we implement a novel "quantumness gate" that maps an arbitrary matrix to the real representation of a positive hermitean normalized density matrix. We train the CNOT gate, the Hadamard gate and a rotation in Hilbert space as basic building blocks for processing the quantum density matrices of two entangled qubits. During the training process the neural networks learn to represent the complex structure, the hermiticity, the normalization and the positivity of the output matrix. The requirement of successful training allows us to find a critical bottleneck dimension which reflects the relevant quantum information. Chains of individually trained neural quantum gates can be constructed to realize any unitary transformation. For scaling to larger quantum systems, we propose to use correlations of stochastic macroscopic two-level observables or classical bits. This novel concept provides a path for a classical implementation of computationally relevant quantum information processing on classical neural networks, in particular on neuromorphic computing machines featuring stochastic operations.
\end{abstract}

\maketitle

\section{\label{sec:level1}Introduction}
Can classical artificial neural networks learn to emulate quantum operations? Does neuromorphic computing have the capabilities for an efficient execution of quantum algorithms? Does our brain use schemes analog to quantum information processing?  A first guess may judge this as unlikely, given that such systems work at room temperature and presumably do not permit well isolated units that can represent a substantial number of ``microscopic'' entangled qubits. This guess is challenged by the observation that quantum operations can be realized within classical statistical systems \cite{wetterich:2018}. In this case quantum states are represented by probabilistic information concerning macroscopic observables. A quantum gate translates to processing this probabilistic information at some ``layer'' $t$ to the neighboring layer $t+\epsilon$ in a particular structured way. Layers may correspond to time, space or levels in neural networks. Consecutive non-commuting quantum operations can be realized by different gates on a sequence of layers. 

In this letter we demonstrate important aspects of quantum computing by small artificial neural networks (ANN). Our network can perform sequences of arbitrary unitary transformations of the density matrix for two entangled qubits. After learning only a set of basic operations such as the Hadamard gate, CNOT-gate and $\pi/8$-rotation, the system can execute chains of such operations. Arbitrary quantum operations for two qubits can be performed by sequences of these basic gates. During the training the network for a particular gate unit learns the complex structure, hermiticity, proper normalization, and positivity of the output density matrix, plus the requested unitary transformation for its assigned task, see Figs. 1,2, with more explanations below. This output matrix can then be processed by further gates that have learned before their particular task by individual training. For arbitrary quantum operations only the training of the basic gates is necessary.

Quantum operations need a processing of probabilistic information at some layer $t$ to probabilistic information at the next layer $t+\epsilon$. This probabilistic information is encoded in the form of a number of expectation values of macroscopic observables. These can include correlations, e.g. expectation values of products of ``basis observables''. For a choice of macroscopic observables it is sufficient to consider two-level observables that can take the values $s=\pm1$. For example, a neuron firing above a given rate may correspond to $s=1$, while firing below this rate is associated to $s=-1$. Such macroscopic observables can be identified with classical Ising spins or classical bits. Another example for a binary neural two-state system is a spiking neuron being either in an active or in the refractory state as discussed in \cite{petrovici:2016}\cite{buesing:2011}.   

Our final goal is the training of a network such that this transformation of probabilistic information realizes quantum operations for a suitable "quantum subsystem" \cite{wetterich:2018}. The quantum subsystem uses only part of the probabilistic information, which is collected in the quantum density matrix $\rho(t)$ at a given layer $t$. This density matrix is the key object of our investigation. For two qubits it is a positive hermitean $4\times 4$-matrix obeying the normalization $\text{tr}\rho=1$. A quantum gate performs a unitary transformation $U(t)$ for obtaining the density matrix at the next neighboring layer at $t+\epsilon$
\be\label{1}
\rho(t+\epsilon)=U(t)\rho(t)U^\dagger(t).
\ee
The elements of the density matrix are associated to suitable combinations of expectation values of macroscopic observables. 

For $Q$ quantum spins or qubits the elements of the density matrix $\rho$ correspond to $2^{2Q}-1$ independent real numbers. Positivity of $\rho$ requires certain quantum constraints \cite{wetterich:2018} for these numbers which, in turn, entail the uncertainty relations of quantum mechanics. If all those numbers are associated to expectation values of independent basis observables, a very large number $2^{2Q}-1$ of independent basis observables would be required. The need of $2^{40}$ basis observables for $20$ entangled qubits would be prohibitive for any practical realization. A drastic change of the scaling properties occurs if elements of $\rho$ are associated to correlations of basis observables. Expectation values and correlations of $3Q$ classical Ising spins can then describe an arbitrary density matrix for $Q$ qubits \cite{wetterich:2018}. Indeed, it has been shown formally that an appropriate processing of the probabilistic information for $3Q$ classical bits is sufficient to realize an arbitrary unitary transformation \eqref{1} for the density matrix for $Q$ qubits. Use of correlations amounts to a tremendous economy in the number of required basis observables. This economy is less dramatic if $\rho$ does not use arbitrarily high correlations. Nevertheless, in view of macroscopic observables often accounting for several classical bits ($2^k$ different states of a macroscopic observable amount to $k$ classical bits), and the huge amount of information stored in correlation functions, the realization of rather large entangled density matrices in terms of expectation values of macroscopic observables seems not unrealistic. Furthermore the full quantum information is often not needed for performing specific tasks.

A central question arises: how to realize the processing of the probabilistic information needed for the unitary transformations \eqref{1}? In this letter we propose that this processing can be performed by ANN. During the training phase the units for individual gates learn implicitly the necessary transformation rule. The investigation of the possibility of quantum computations by artificial neural networks (ANN) can be split into several steps. In a first step one takes the expectation values of the observables from which $\rho$ is constructed as real numbers, without reference to a probabilistic origin. At this level no correlations are used, and the elements of the density matrix can be considered equivalently as expectation values of observables or classical values of some continuous observables. The demonstration of this step for two qubits is the purpose of this letter.

A second step beyond the present letter should associate these real numbers to expectation values and correlations of suitable macroscopic basis observables. Further steps have to investigate the scaling with an increasing number of qubits, questions of errors, and finally a practical emulation of probabilistic quantum computing by ANN or special hardware like neuromorphic computers. In particular the latter offer attractive features like massively parallel processing,  computing with accelerated operation and on-chip learning capabilities (see e.g. \cite{aamir:2017} and \cite{friedmann:2017} for recent implementations of physical model neuromorphic systems). Neuromorphic systems also provide the attractive feature of self-generated stochasticity \cite{dold:2018}.  

Recent work \cite{killoran:2018} describes a quantum neural network model as variational quantum circuits built in a continuous-variable architecture. They developed a software framework for optimizing and evaluating optical quantum circuits, with a tensorflow backend as a simulator. OpenFermion \cite{mcclean:2017} is an open-source software library for the simulation of fermionic models and quantum chemistry problems. The Cirq project \cite{google:2018} is a simulator for near term quantum devices with tensorflow and a Google hardware backend. Neural networks have been proposed in the past for performing particular tasks necessary for quantum computation \cite{swaddle:2017} or representing aspects of quantum systems \cite{carleo:2016}. In this letter we propose to use ANN directly for performing operations of quantum computing.
\section{\label{sec:level1}Implementation}
For the implementation of a given gate for two qubits, i.e. a given unitary transformation \eqref{1}, we consider a network whose input is a real $8\times 8$-matrix $A(t)$, and the output a real $8\times 8$-matrix $B(t+\epsilon)$. These matrices are encoded in $64$ input ``neurons'', representing the matrix elements, and correspondingly $64$ output neurons, all realized as real linear units. We employ one intermediate "bottleneck" layer with $m$ neurons, corresponding to a network structure $64-m-64$. The quantum subsystem characterized by the density matrix $\rho$ involves only $15$ independent numbers, and the network has to learn that only those are important. The remaining $64-15=49$ numbers can be considered as the "probabilistic environment" of the quantum subsystem. The information concerning this environment is not available to the quantum subsystem.

We first describe a "quantumness gate" that maps an arbitrary matrix $A(t)$ to the real representation of a positive hermitean normalized density matrix $\rho(t)$ that characterizes the quantum subsystem. We draw the $64$ matrix elements of $A(t)$ from a uniform distribution between $1$ and $-1$.  After applying the gate all quantum conditions and uncertainty relations are realized. The map is done in two steps. The first maps $A$ into a complex $4\times 4$ matrix $\hat C$. For this purpose we employ
\be\label{2}
\tilde A=-IAI~,~
I= \begin{pmatrix}
   0 & -1 \\
   1 & 0 
\end{pmatrix},
\ee
with $1$ denoting the $4\times 4$ unit matrix. The matrix $\bar A=\frac12(A+\tilde A)$ can be written as
\be\label{3}
\bar A=1_2 \otimes C_R+I_2\otimes C_I=
\begin{pmatrix}
   C_R-C_I \\
   C_I~~~C_R
\end{pmatrix},
\ee
with real $4\times 4$-matrices $C_R$ and $C_I$. The complex $4\times 4$-matrix $\hat C$ is defined by $\hat C=C_R+iC_I$. (We always denote by subscripts $R$ and $I$ the real and imaginary parts of a complex matrix.) This construction is compatible with complex matrix multiplication. The real matrix product $\bar A_1\bar A_2$ is mapped to the complex matrix product $\hat C_1\hat C_2$. We call $\bar A$ the real representation of the complex matrix $\hat C$. The second step constructs from $\hat C(t)$ the positive hermitean normalized density matrix $\rho(t)$ as
\be\label{4}
\rho(t)=\frac{\hat C(t)\hat C^\dagger(t)}{{\rm tr }{\Big\{ \hat C(t)\hat C^\dagger (t)\Big\}}}.
\ee
The real representation of the density matrix $\bar{\rho}$ is formed as in equation \eqref{3}. 

The combination of these steps in a "quantumness gate" maps an arbitrary matrix $A(t)$ to an associated real representation of the density matrix $\bar{\rho}(t+\epsilon)$. We have constructed an ANN that realizes the quantumness gate so far with a precision of $4 \times 10^{-5}$. This network uses rectify linear non-linearities, the AdaDelta optimizer \cite{zeiler:2012} for training and increasing batch sizes of $32,64,\ldots,512$ on $10^6$ training samples. All ANNs are implemeted in Tensorflow \cite{tensorflow:2015} using the Keras \cite{chollet:2015} library.

If we start with fixed values of Ising spins, as encoded by all elements of the input matrix $A(t)$ being either one or minus one, the output density matrix $B(t+\epsilon)$ enforces the uncertainty relations of quantum mechanics, such that not all elements of $B(t+\epsilon)$ can have the values $\pm 1$.  In short, the network learns that all essential information is stored in the quantum subsystem characterized by $\rho(t)$ by minimizing the difference between the output matrix $B(t+\epsilon)$ and the real density matrix $\bar{\rho}$ constructed by equations \eqref{2}-\eqref{4}.

For the training of the quantum gates we use real representations of density matrices as input matrices. We want the ANN to learn that the output matrix is the real representation of the unitary transformation of the input density matrix. For this purpose the loss $C$ to be minimized by the learning is the mean squared error between $B(t+\epsilon)$ and $\bar \rho(t+\epsilon)$
\be\label{5}
C= \frac{1}{N} \sum^N_{i=0} ||B_i(t+\epsilon)-\bar \rho_i(t+\epsilon)||^2
\ee
where $i$ indexes the training samples and $\bar\rho(t+\epsilon)$ is the real representation of $\rho(t+\epsilon)$, similar to eq. \eqref{3},
\be\label{5A}
\bar \rho(t+\epsilon)=1_2\otimes \rho_R(t+\epsilon)+I_2\otimes ~\rho_I(t+\epsilon).
\ee
Here $\rho(t+\epsilon)$ is given by eq. \eqref{1} with $U(t)$ the unitary transformation to be learned, for example the CNOT-gate. In equation \eqref{5} the norm $||~||$ is the Frobenius norm of complex or real matrices. The Adagrad optimizer \cite{duchi:2010} is used for training, with $N=10^5$ training samples and a batch size of $10^3$. Without any training the output matrix $B(t+\epsilon)$ is an arbitrary real matrix. After the training the network has learned the complex structure in the sense that the matrix $B(t+\epsilon)$ is the real representation of a complex $4\times 4$-matrix $\hat F(t+\epsilon)$, given by 
\be\label{6}
B(t+\epsilon)=
\left(
\begin{array}{lr}
\hat F_R(t+\epsilon),&-\hat F_I(t+\epsilon)\\ 
\hat F_I(t+\epsilon),&\hat F_R(t+\epsilon)
\end{array}\right).
\ee
Furthermore, it has learned the hermiticity of $\hat F(t+\epsilon)$ and the normalization tr$\hat F_R(t+\epsilon)=\frac12$tr$B(t+\epsilon)=1$, as well as the positivity of $\hat F(t+\epsilon)$. (All eigenvalues of $\hat F(t+\epsilon)$ are positive semi-definite.) Finally, it has learned that $\hat F(t+\epsilon)$ is given by the wanted unitary transformation of $\rho(t)$ according to eq. \eqref{1}. After the training the parameters of the network have adapted such that an arbitrary input density matrix  $\rho(t)$ is transformed to the real representation of the density matrix $\bar \rho(t+\epsilon)$, with $\rho(t+\epsilon)$ the unitary transform \eqref{1} of the density matrix $\rho(t)$.  

\begin{figure}
\includegraphics{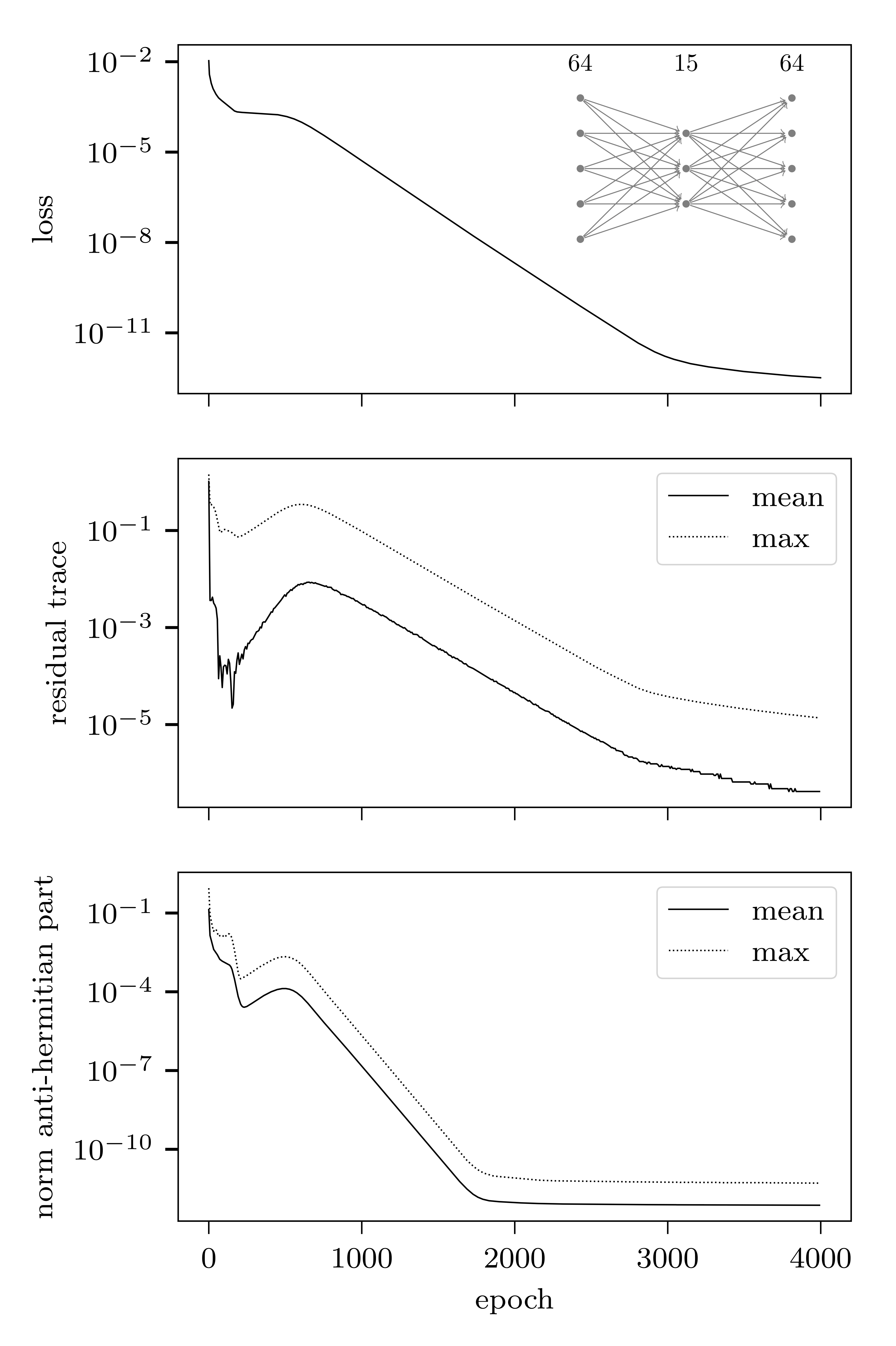}
\caption{\label{figure:1} Loss function (upper plot), residual trace (distance of the trace from $1$) (middle plot) and norm of the anti-hermitian part (lower plot) of the output density matrix for supervised training on the CNOT gate. The dimension of the bottleneck layer is chosen to be $m = 15$.}
\end{figure}

\section{\label{sec:level1}Results}

In Fig. \ref{figure:1} we show the loss (i.e. the value of the cost function \eqref{5}) for the CNOT-gate as a function of the number of training steps (epochs) for the number of intermediate neurons $m=15$. Fig. 1 also displays the normalization $||\text{tr}\hat F(t+\epsilon)-1||$ and the norm of the antihermitean part $||\hat F(t+\epsilon)-\hat F^\dagger(t+\epsilon)||$. Already after a short initial period the network has learned the norm and hermiticity of the output matrix $\hat F(t+\epsilon)$ to a good precision. This holds both for an individual training ( displayed here for the maximum deviation from the desired value) and for the mean of $10^5$ independent initial matrices. In Fig. 2 we display the loss function for different numbers of intermediate neurons $m$. For $m$ above $16$ the learning saturates already after a modest number of $1000$ steps. The critical number $m=15$, for which the learning is somewhat slower, corresponds precisely to the number of independent elements of a normalized hermitean $4 \times 4$ matrix. For $m$ smaller than $15$ the learning success breaks down dramatically.

We can therefore use the structure of our network for a determination of the minimal information needed to achieve the assigned task. This procedure may be generalized to settings where the needed information is less obvious as for the present investigation. One could investigate in this way if really a quantum density matrix is employed by the network for a given task. Replacing for the quantum gate the input matrix $\bar\rho(t)$ by an arbitrary matrix $A(t)$, the trained network will, in general, not produce a normalized hermitean output matrix. An exception is $m=15$.

\begin{figure}
\includegraphics{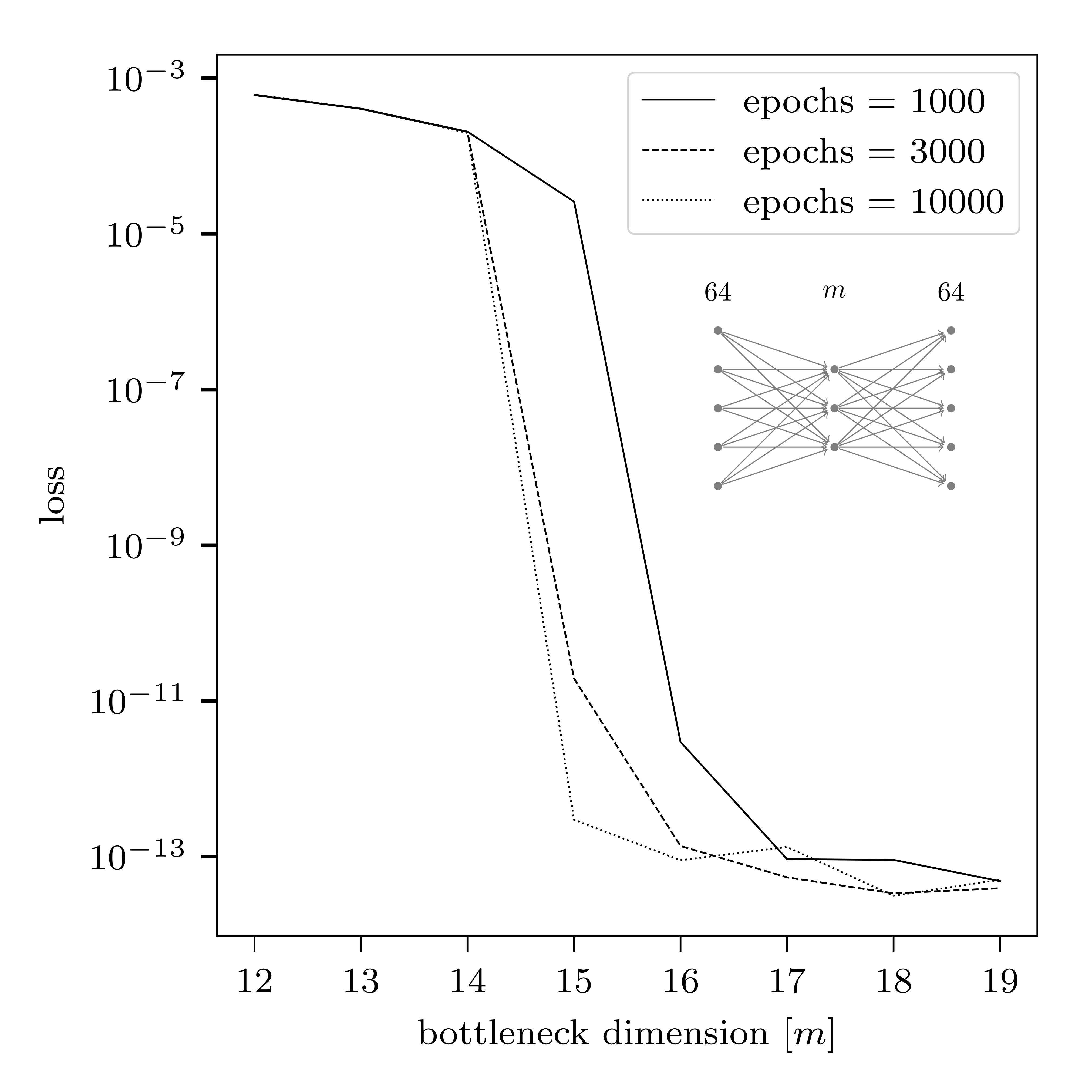}
\caption{\label{figure:4} Loss function value after $500$, $1000$ and $3000$ epochs of training on the CNOT gate. The network learns successfully provided the number of intermediate neurons is $m \geq 15$, corresponding to the number of independent elements of a normalized hermitean $4 \times 4$ matrix.}
\end{figure}

We next establish the performance after the training phase. We first train separately two units, one corresponding to the CNOT gate $U_C$, and the second $(U_{HR})$ to the Hadamard gate for the first quantum spin, combined with a $\pi/8$ rotation for the second quantum spin. The two gates are given explicitly by 
\ba\label{7}
U_C&=&\begin{pmatrix}
       1&0\\0&\tau_1
      \end{pmatrix},~
U_{HR}=U_{H1}U_{R2},\nn\\
U_{H1}&=&\frac{1}{\sqrt{2}}
\begin{pmatrix}
 1&1\\1&-1
\end{pmatrix},
U_{R2}=
\begin{pmatrix}
U_T&0\\0&U_T 
\end{pmatrix},
\ea
with $2\times 2$ sub-matrices
\be\label{8}
\tau_1=\begin{pmatrix}
        0&1\\1&0
       \end{pmatrix},
U_T=
\begin{pmatrix}
1&0\\0&~~~e^{i\pi/4} 
\end{pmatrix}.
\ee
Each individual gate is trained separately with $3000$ steps.

\begin{figure}
\includegraphics{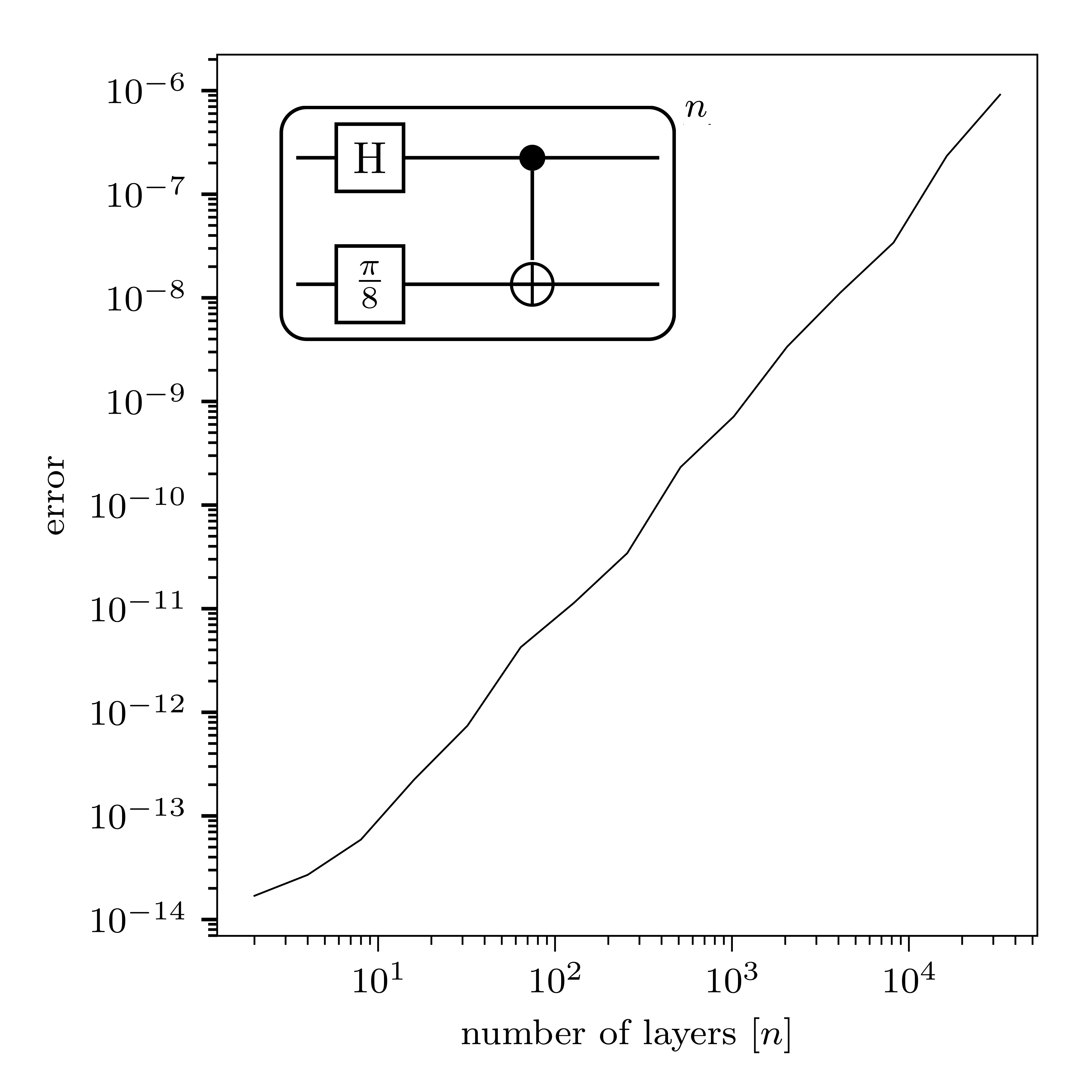}
\caption{\label{figure:3} Repeated execution of the gate $\bar U$ shown in the inset in $n$ layers. We display the mean squared error after composing the circuit for a given number of layers.}
\end{figure}

For a sequence of two gates we use the output matrix $B(t+\epsilon)$ as input matrix for the second gate. In other words, we replace $\bar\rho(t)$ by $B(t+\epsilon)$. We perform in this way a sequence of $n$ combined gates $\bar U=U_CU_{HR}$, without any further feedback. We compare the output matrix $B(t+n\epsilon)$ with a result of a sequence of unitary transformations $\rho(t+n\epsilon)$,
\ba\label{8a}
\rho(t+n\epsilon)&=&\tilde U(t+n\epsilon,t)\rho(t)\tilde U^\dagger(t+n\epsilon,t),\\
\tilde U(t+n\epsilon,t)&=&\bar U\big(t+(n-1)\epsilon\big)\dots \bar U(t+\epsilon)\bar U(t)=\bar U(t)^n.\nn
\ea
We plot in Fig. \ref{figure:3} the mean square error after a certain number of layers $n$. We observe that $B(t+n\epsilon)$ approximates $\bar\rho(t+ n\epsilon)$ rather well for $n$ as large as $2^{15}$. We emphasize that the products for various n realize a very dense set of unitary transformations, demonstrating that arbitrary transformations can be realized, without the need of explicitly training corresponding gates. The non-commutativity of the quantum gates can easily be demonstrated by changing the order of $U_C$ and $U_{HR}$ in the definition of $\bar U$, e.g. $\bar U'=U_{HR}U_C$. This is done by changing the order of the gates in the sequence.

\section{\label{sec:level1}Summary and Outlook}

In this letter we have discussed general concepts for the emulation of quantum computation by artificial neural networks. We have successfully implemented neural networks that realize arbitrary quantum operations for two qubits. In the present investigation the probabilistic nature of quantum computing is not yet manifest, since we operate with real numbers representing expectation values, but not explicitly with expectation values and correlations in stochastic systems. In the future we plan to work with a larger number of neurons and define the elements of the matrices $A$ and $B$ as expectation values of observables that are formed by averaging over all or a subset of neurons of a given layer. An interesting alternative are time averages over variables changing rapidly according to stochastic evolution equations. The network may learn to update the parameters of the stochastic equations, thereby modifying the expectation values. The learning goals can be defined with the same prescriptions for density matrices as in the present work. We expect similar results at least for the case of a few qubits. Realizing an efficient scaling for a larger number of qubits by exploiting correlation functions will be an interesting challenge and will reveal what kind of resources are necessary for a given task. If met, the way to quantum computing by neural networks or neuromorphic computing may indeed be open.

\section{\label{sec:level1}Acknowledgements}

This work has been initiated by the preparatory work for the DFG Excellence Cluster STRUCTURES. It was supported by the EU Horizon 2020 framework program under grant agreement 720270 (Human Brain Project), the DFG Collaborative Research Centre SFB1225 (ISOQUANT), by the  ERC Advanced Grant EntangleGen Project-ID 694561 and the Heidelberg Graduate School for Fundamental Physics (HGSFP).
\section{\label{sec:level1}References}

\bibliography{Quantum_computation_with_neural_networks}
\end{document}